\documentclass[a4paper,10pt,reqno]{amsart}
\usepackage{a4wide}
\usepackage{amsfonts,amsmath,amssymb}
\usepackage{graphicx}
\usepackage{tikz}

\theoremstyle{remark}

\def\be{\begin{equation}}
\def\ee{\end{equation}}
\def\bea{\begin{eqnarray}}
\def\eea{\end{eqnarray}}

\def\mtr{\operatorname{tr}}
\def\adj{\operatorname{adj}}

\def\scq{{\textsc q}}

\begin{document}

\thispagestyle{plain}

\title{On the algebraic area of lattice walks and the Hofstadter model}

\author{St\'ephane Ouvry (*), Stephan Wagner (**) and Shuang Wu (*)}

\date{\today}

\begin{abstract}
We consider the generating function of the algebraic area of lattice walks, evaluated at a root of unity, and its relation to the Hofstadter model. In particular, we obtain an expression for the generating function of the $n$-th moments of the Hofstadter Hamiltonian in terms of a complete elliptic integral, evaluated at a rational function. This in turn gives us both exact and asymptotic formulas for these moments.
\end{abstract}

\maketitle

(*) LPTMS, CNRS-Facult\'e des Sciences d'Orsay, Universit\'e Paris Sud, 91405 Orsay Cedex, France

(**)  Department of Mathematical Sciences, Stellenbosch University, Matieland 7602, South Africa
\section{Introduction}

The algebraic area is the area enclosed by a curve, weighted by the winding number: if the curve moves around a region in counterclockwise (positive) direction, its area counts as positive, otherwise negative. Moreover, if the curve winds around more than once, the area is counted with multiplicity. In this paper, we will be studying the algebraic area of two-dimensional lattice walks starting at the origin and moving up, down, left or right at each step. If the walk is not closed, we define its algebraic area as that of the closed walk obtained by connecting its endpoint with its starting point, adding on to the end of the walk the minimum necessary number of steps, first vertical, then horizontal.

\begin{figure}[htbp]
\begin{center}
\begin{tikzpicture}[scale=0.6]
\draw [dashed, fill=lightgray] (0,0)--(0,-2)--(3,-2)--(3,1)--(2,1)--(2,2)--(1,2)--(1,0)--(0,0);
\draw [very thick, fill=lightgray] (3,1)--(5,1)--(5,2)--(4,2)--(4,3)--(3,3)--(3,1);
\draw [very thick] (0,0)--(0,-2)--(3,-2)--(3,1)--(2,1)--(2,2)--(1,2);
\draw [very thick, fill=lightgray] (5,0)--(6,0)--(6,1)--(5,1)--(5,0);
\node[fill=black,circle,inner sep=2pt] () at (0,0) {};
\node at (1.5,-0.5) {$+$};
\node at (3.5,1.5) {$-$};
\node at (5.5,0.5) {$+$};
\end{tikzpicture}
\end{center}
\caption{A lattice walk with algebraic area $9-3+1$.}
\end{figure}
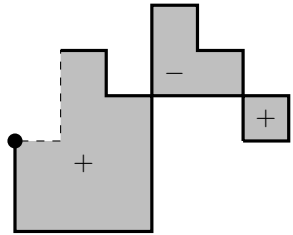

Suppose that a lattice walk moves $m_1$ steps right, $m_2$ steps left, $l_1$ steps up and $l_2$ steps down. If e.g. $m_1 \ge  m_2$ and $l_1 \ge l_2$, 
we add $l_1 - l_2$ steps down followed by $m_1 - m_2$ steps to the left in order to close the path. Let $C_{m_1,m_2,l_1,l_2}(A)$ be the number of such walks enclosing  an algebraic area $A$. Finding the generating function for the $C_{m_1,m_2,l_1,l_2}(A)$'s, i.e.,
\be Z_{m_1,m_2,l_1,l_2}(\scq)=\sum_A C_{m_1,m_2,l_1,l_2}(A) \scq^A \label{z_defi},\ee 
is a difficult task so far without a definite answer\footnote{For recent efforts in this direction see e.g. \cite{epelbaum,desbois}.}. 

One can  simplify  by restricting to closed lattice walks of a given length $n$ ($n$ is then necessarily even), i.e., walks with an equal number $m$ of steps right/left and an equal number $n/2-m$ of steps up/down,  with $m \in \{0,1,\ldots,n/2\}$,   and by   focusing on the generating function  
$$Z_n(e^{i\gamma})=\sum_{m=0}^{n/2}Z_{m, m, \frac{n}{2}-m, \frac{n}{2}-m}(e^{i\gamma})$$
evaluated at $\scq=e^{i\gamma}$, a root of unity.
  One reason for addressing this  simpler question  arises from  the deep connection  between the algebraic area distribution of random walks  to the quantum mechanics of a charged particle in a perpendicular magnetic field. A paramount example is, in the continuous limit, Levy's law \cite{levy} for the algebraic area distribution of closed Brownian curves and its connection to the quantum Landau problem.
 
 In the discrete case,  the connection  is to the quantum Hofstadter model \cite{Hofstadter} of a particle hopping on a two-dimensional lattice  in  a rational magnetic flux $\gamma=2\pi p/q$, here counted in unit of the flux quantum. More precisely, $Z_n(e^{i\gamma})$ is mapped \cite{bellissard} to the $n$-th moment of the Hofstadter Hamiltonian $H_{\gamma}$:
\be \label{map}Z_{n}(e^{i\gamma})={\rm Tr}\:H_{\gamma}^{n}.\ee

In the following section, we will derive an expression for the multivariate generating function of $Z_{m_1,m_2,l_1,l_2}$ as defined in~\eqref{z_defi}, which even yields simple explicit formulas in some very special cases. This multivariate generating function is then specialised to the generating function of the $Z_n(e^{i\gamma})$'s.
This allows us to derive a closed expression for the $Z_n(e^{i\gamma})$'s und thus, by virtue of \eqref{map}, for the traces ${\rm Tr}\:H_{\gamma}^{n}$, in terms of the Kreft coefficients \cite{Kreft}.  Not surprisingly, these coefficients     encode in a simple way  the  quantum  secular equation determining the Hofstadter spectrum. We will also use the generating function to study the asymptotic behaviour of $Z_n(e^{i\gamma})$ as $n \to \infty$ for fixed $\gamma$.

\section{The generating function for arbitrary walks}
 We are interested in evaluating the generating function  $Z_{m_1,m_2,l_1,l_2}(\scq)$ when $\scq$ is a root of unity. Of course, we trivially have
$$Z_{m_1,m_2,l_1,l_2}(1) = \frac{(m_1+m_2+l_1+l_2)!}{m_1!m_2!l_1!l_2!}.$$

\medskip

Let us start with some symmetry properties: first of all, it is easy to see that
\begin{equation}\label{eq:refl1}
Z_{m_2,m_1,l_1,l_2}(\scq) = Z_{m_1,m_2,l_2,l_1}(\scq) = Z_{m_1,m_2,l_1,l_2}(\scq^{-1}),
\end{equation}
since reflections about the $x$- or $y$-axis only change the sign of the algebraic area. Moreover,
\begin{equation}\label{eq:refl2}
Z_{m_1,m_2,l_1,l_2}(\scq) = Z_{l_1,l_2,m_1,m_2}(\scq).
\end{equation}
Finally, if we reverse the direction of a walk with $m_1,m_2,l_1,l_2$ steps right, left, up and down, respectively, the algebraic areas of the original and the reversed walk add up to $-(m_1-m_2)(l_1-l_2)$, which is particularly evident if the walk only moves in one quadrant, as in Figure~\ref{fig:adding}.

\begin{figure}[htbp]
\begin{center}
\begin{tikzpicture}[scale=0.6]
\draw [dashed, fill=lightgray] (0,0)--(0,1)--(1,1)--(1,3)--(2,3)--(2,4)--(5,4)--(5,0)--(0,0);
\draw [very thick] (0,0)--(0,1)--(1,1)--(1,3)--(2,3)--(2,4)--(5,4);
\draw [dashed, fill=lightgray] (0,0)--(-3,0)--(-3,-1)--(-4,-1)--(-4,-3)--(-5,-3)--(-5,0)--(0,0);
\draw [very thick] (0,0)--(-3,0)--(-3,-1)--(-4,-1)--(-4,-3)--(-5,-3)--(-5,-4);
\node[fill=black,circle,inner sep=2pt] () at (0,0) {};
\end{tikzpicture}
\end{center}
\label{fig:adding}
\caption{Illustration of~\eqref{eq:adding}.}
\end{figure}

It follows that
\begin{equation}\label{eq:adding}
Z_{m_2,m_1,l_2,l_1}(\scq) = \scq^{-(m_1-m_2)(l_1-l_2)} Z_{m_1,m_2,l_1,l_2}(\scq^{-1}),
\end{equation}
and combining this with~\eqref{eq:refl1} yields
$$Z_{m_1,m_2,l_1,l_2}(\scq) = \scq^{-(m_1-m_2)(l_1-l_2)} Z_{m_1,m_2,l_1,l_2}(\scq^{-1}).$$
Thus
$$\scq^{(m_1-m_2)(l_1-l_2)/2}Z_{m_1,m_2,l_1,l_2}(\scq)$$
is symmetric (in the sense that the coefficients of $\scq^k$ and $\scq^{-k}$ are equal for all $k$). If $|\scq| = 1$ (in particular, if $\scq$ is a root of unity), this implies that
$$\scq^{(m_1-m_2)(l_1-l_2)/2}Z_{m_1,m_2,l_1,l_2}(\scq)$$
is real.

For further study, we define the multivariate generating function
\be
A(x_1,x_2,y_1,y_2,\scq) = \sum_{m_1,m_2,l_1,l_2 \geq 0} Z_{m_1,m_2,l_1,l_2}(\scq) x_1^{m_1}x_2^{m_2} y_1^{l_1} y_2^{l_2}.\label{A_defi}
\ee
We distinguish the four possible cases for the last step: if the last step is vertical, then the algebraic area does not change. If it is a step to the left, the algebraic area changes exactly by the current $y$-coordinate; the same holds if the last step is a step to the right, but with opposite sign. It is now easy to see that the recursion
\begin{align*}
Z_{m_1,m_2,l_1,l_2}(\scq) &= Z_{m_1,m_2,l_1-1,l_2}(\scq) + Z_{m_1,m_2,l_1,l_2-1}(\scq) \\
&\quad +\scq^{l_2-l_1} Z_{m_1-1,m_2,l_1,l_2}(\scq) + \scq^{l_1-l_2} Z_{m_1,m_2-1,l_1,l_2}(\scq)
\end{align*}
holds, with initial values $Z_{0,0,0,0}(\scq) = 1$ and $Z_{m_1,m_2,l_1,l_2}(\scq) = 0$ whenever $\min(m_1,m_2,l_1,l_2) < 0$. We immediately obtain a functional equation for the multivariate generating function:
\begin{align*}
A(x_1,x_2,y_1,y_2,\scq) &= 1 + y_1 A(x_1,x_2,y_1,y_2,\scq) + y_2 A(x_1,x_2,y_1,y_2,\scq) \\
&\quad + x_1 A(x_1,x_2,\scq^{-1}y_1,\scq y_2,\scq) + x_2 A(x_1,x_2,\scq y_1,\scq^{-1}y_2,\scq).
\end{align*}
When $\scq$ is a root of unity, this turns into a finite system of linear equations. In the following, we will assume that $\scq$ is a (without loss of generality primitive) $q$-th root of unity, so that $\scq^q = 1$.
Set
$$A_k = A(x_1,x_2,\scq^ky_1,\scq^{-k}y_2,\scq),$$
which only depends on the residue class of $k$ modulo $q$. Note that in particular $A_0 = A_q = A(x_1,x_2,y_1,y_2,\scq)$.
Now we have
\begin{align*}
A_k &= A(x_1,x_2,\scq^ky_1,\scq^{-k}y_2,\scq) \\
&= 1 + \scq^ky_1 A(x_1,x_2,\scq^ky_1,\scq^{-k}y_2,\scq) + \scq^{-k}y_2 A(x_1,x_2,\scq^ky_1,\scq^{-k}y_2,\scq) \\
&\quad + x_1 A(x_1,x_2,\scq^{k-1}y_1,\scq^{-k+1}y_2,\scq) + x_2 A(x_1,x_2,\scq^{k+1}y_1,\scq^{-k-1}y_2,\scq) \\
&= 1 + (\scq^ky_1+\scq^{-k}y_2)A_k + x_1 A_{k-1} + x_2 A_{k+1}.
\end{align*}
The resulting linear system can be written in matrix form as
$$\begin{pmatrix}
c_0  & -x_2 & 0 & \cdots & 0 & -x_1 \\
-x_1 & c_1  & -x_2 & \cdots & 0 & 0 \\
0 & -x_1 & c_2  & \cdots & 0 & 0 \\
\vdots & \vdots & \vdots & \ddots & \vdots & \vdots \\
0 & 0 & 0  & \cdots & c_{q-2} & -x_2 \\
-x_2 & 0 & 0  & \cdots & -x_1 & c_{q-1} \\
\end{pmatrix} \cdot \begin{pmatrix} A_0 \\ A_1 \\ A_2 \\ \vdots \\ A_{q-2} \\ A_{q-1} \end{pmatrix} = \begin{pmatrix} 1 \\ 1 \\ 1 \\ \vdots \\ 1 \\ 1 \end{pmatrix},$$
where $c_k = 1 - \scq^ky_1 - \scq^{-k}y_2$. We can apply Cramer's rule to solve for $A_0$, which immediately shows that it is a rational function in $x_1,x_2,y_1,y_2$. 

\subsection{Special cases}

For small values of $q$, the generating function is simple enough to extract explicit formulas for $Z_{m_1,m_2,l_1,l_2}(\scq)$ from it.

\begin{itemize}
\item Specifically, we obtain for $q = 2$ (and thus $\scq = -1$)
$$A_0 = A(x_1,x_2,y_1,y_2,-1) = \frac{1+x_1+x_2+y_1+y_2}{1-(x_1+x_2)^2-(y_1+y_2)^2}.$$
Expanding into a power series, we find
\begin{align*}
A(x_1,x_2,y_1,y_2,-1) &= \frac{1+x_1+x_2+y_1+y_2}{1-(x_1+x_2)^2-(y_1+y_2)^2} \\
&= (1+x_1+x_2+y_1+y_2) \sum_{n \geq 0} ((x_1+x_2)^2+(y_1+y_2)^2)^n \\
&= (1+x_1+x_2+y_1+y_2) \sum_{n_1 \geq 0} \sum_{n_2 \geq 0} \binom{n_1+n_2}{n_1}(x_1+x_2)^{2n_1}(y_1+y_2)^{2n_2} \\
&= \sum_{n_1 \geq 0} \sum_{n_2 \geq 0} \binom{n_1+n_2}{n_1} \Big( (x_1+x_2)^{2n_1}(y_1+y_2)^{2n_2} + (x_1+x_2)^{2n_1+1}(y_1+y_2)^{2n_2} \\
&\quad + (x_1+x_2)^{2n_1}(y_1+y_2)^{2n_2+1} \Big).
\end{align*}
From this expression, one reads off the coefficients easily:
$$Z_{m_1,m_2,l_1,l_2}(-1) = \begin{cases} \binom{\lfloor (m_1+m_2+l_1+l_2)/2 \rfloor}{ \lfloor (m_1+m_2)/2 \rfloor } \binom{m_1+m_2}{m_1} \binom{l_1+l_2}{l_1} & \text{if } m_1+m_2 \text{ or } l_1+l_2 \text{ or both are even,} \\
0 & \text{otherwise.}\end{cases}$$
\item In general, however, the generating function $A$ becomes more complicated. For $q=3$, we already obtain
$$\frac{1+x_1+x_2+y_1+y_2+x_1^2+x_2^2+y_1^2+y_2^2 - x_1x_2-y_1y_2-\scq x_1y_1-\scq^{-1}x_1y_2-\scq^{-1}x_2y_1-\scq x_2y_2}{1-x_1^3-x_2^3-y_1^3-y_2^3-3x_1x_2-3y_1y_2},$$
which gives us, for example in the case where $m_1 - m_2 \equiv l_1 - l_2 \equiv 1 \bmod 3$,
$$Z_{m_1,m_2,l_1,l_2}(e^{2\pi i/3}) = -e^{2\pi i/3} \sum_{\substack{k = 0 \\ k \equiv m_2 \bmod 3}}^{\min(m_1,m_2)} \sum_{\substack{j = 0 \\ j \equiv l_2 \bmod 3}}^{\min(l_1,l_2)}
3^{k+j} \binom{\frac{m_1+m_2+l_1+l_2+k+j-2}{3}}{k,j, \frac{m_1-k-1}{3}, \frac{m_2-k}{3}, \frac{l_1-j-1}{3}, \frac{l_2-j}{3}}.$$
Similar formulas hold when $m_1-m_2$ and $l_1-l_2$ lie in other fixed residue classes modulo $3$.
\item An elegant formula is obtained for $\scq = \pm i$ (i.e., $q=4$) if we count walks by the total number of horizontal and vertical steps: we have

$$\sum_{m_1+m_2 = m} \sum_{l_1+l_2 = l} Z_{m_1,m_2,l_1,l_2}(\pm i) = 2^{m+l} \binom{ \lfloor m/2 \rfloor + \lfloor l/2 \rfloor}{\lfloor m/2 \rfloor},$$
which is obtained from the generating function

$$A(x,x,y,y,\pm i) = \frac{1+2x+2y+4xy}{1-4x^2-4y^2}.$$

Note that we also have

$$\sum_{m_1+m_2 = m} \sum_{l_1+l_2 = l} Z_{m_1,m_2,l_1,l_2}(-1) = 2^{m+l} \binom{ \lfloor m/2 \rfloor + \lfloor l/2 \rfloor}{\lfloor m/2 \rfloor},$$
except when $m,l$ are both odd, in which case the sum evaluates to $0$.
\end{itemize}

\subsection{The general shape}

We know now that $A$ is a rational function if evaluated at a root of unity $\scq$. Let us show that it always has the following shape:
\be
A(x_1,x_2,y_1,y_2,\scq) = \frac{U(x_1,x_2,y_1,y_2)}{1 - x_1^q - x_2^q - y_1^q - y_2^q + V(x_1x_2,y_1y_2)},\label{gen_shape}
\ee
where $U,V$ are polynomials and $V$ is symmetric (in its two variables) and has only real coefficients (and constant coefficient $0$). By Cramer's rule, we have
\be A(x_1,x_2,y_1,y_2,\scq) = \frac{\begin{vmatrix}
1  & -x_2 & 0 & \cdots & 0 & -x_1 \\
1 & c_1  & -x_2 & \cdots & 0 & 0 \\
1 &-x_1 & c_2  & \cdots & 0 & 0 \\
\vdots & \vdots & \vdots & \ddots & \vdots & \vdots \\
1 & 0 & 0  & \cdots & c_{q-2} & -x_2 \\
1 & 0 & 0  & \cdots & -x_1 & c_{q-1} \\
\end{vmatrix}}{\begin{vmatrix}
c_0  & -x_2 & 0 & \cdots & 0 & -x_1 \\
-x_1 & c_1  & -x_2 & \cdots & 0 & 0 \\
0 & -x_1 & c_2  & \cdots & 0 & 0 \\
\vdots & \vdots & \vdots & \ddots & \vdots & \vdots \\
0 & 0 & 0  & \cdots & c_{q-2} & -x_2 \\
-x_2 & 0 & 0  & \cdots & -x_1 & c_{q-1} \\
\end{vmatrix}},\label{Cramer}\ee
where $c_k = 1 - \scq^ky_1 - \scq^{-k}y_2$ as before. Let us show that the determinant $\Delta$ which appears in  the denominator of \eqref{Cramer} is indeed of the form $1-x_1^q-x_2^q-y_1^q-y_2^q+V(x_1x_2,y_1y_2)$, where $V$ is symmetric and has real coefficients. This is done in the following steps:

\begin{itemize}
\item We note that the determinant $\Delta$ is a polynomial in $x_1,x_2,y_1,y_2$ with total degree $q$.
\item
If we replace $y_1$ by $\scq y_1$ and $y_2$ by $\scq^{-1}y_2$, we end up with the determinant
$$\begin{vmatrix}
c_1  & -x_2 & 0 & \cdots & 0 & -x_1 \\
-x_1 & c_2  & -x_2 & \cdots & 0 & 0 \\
0 & -x_1 & c_3  & \cdots & 0 & 0 \\
\vdots & \vdots & \vdots & \ddots & \vdots & \vdots \\
0 & 0 & 0  & \cdots & c_{q-1} & -x_2 \\
-x_2 & 0 & 0  & \cdots & -x_1 & c_0 \\
\end{vmatrix},$$
which is also obtained from the original determinant $\Delta$ by moving the first row and column to the end; hence, the value does not change. This means that the only nonzero terms in $\Delta$ (seen as a polynomial in $y_1$ and $y_2$ only) that can have a nonzero coefficient are $y_1^q$, $y_2^q$ and all powers of $y_1y_2$.
\item
By expansion with respect to the last row and the last column, we obtain
\begin{align*}
\begin{vmatrix}
c_0  & -x_2 & 0 & \cdots & 0 & -x_1 \\
-x_1 & c_1  & -x_2 & \cdots & 0 & 0 \\
0 & -x_1 & c_2  & \cdots & 0 & 0 \\
\vdots & \vdots & \vdots & \ddots & \vdots & \vdots \\
0 & 0 & 0  & \cdots & c_{q-2} & -x_2 \\
-x_2 & 0 & 0  & \cdots & -x_1 & c_{q-1} \\
\end{vmatrix} = -x_1^q - x_2^q + c_{q-1} \begin{vmatrix}
c_0  & -x_2 & 0 & \cdots & 0 & \\
-x_1 & c_1  & -x_2 & \cdots & 0 \\
0 & -x_1 & c_2  & \cdots & 0  \\
\vdots & \vdots & \vdots & \ddots & \vdots \\
0 & 0 & 0  & \cdots & c_{q-2} \\
\end{vmatrix} \\
-x_1x_2 \begin{vmatrix}
c_0  & -x_2 & 0 & \cdots & 0 & \\
-x_1 & c_1  & -x_2 & \cdots & 0 \\
0 & -x_1 & c_2  & \cdots & 0  \\
\vdots & \vdots & \vdots & \ddots & \vdots \\
0 & 0 & 0  & \cdots & c_{q-3} \\
\end{vmatrix} - x_1x_2 \begin{vmatrix}
c_1  & -x_2 & \cdots & 0 & 0 \\
-x_1 & c_2  & \cdots & 0 & 0 \\
\vdots & \vdots & \ddots & \vdots & \vdots \\
0 & 0 & \cdots & c_{q-3} & -x_2 \\
0 & 0 & \cdots & -x_1 & c_{q-2} \\
\end{vmatrix}.
\end{align*}
It is easy to see that the determinants
\be
D_k = \begin{vmatrix}
u_1  & -x_2 & \cdots & 0 & 0 \\
-x_1 & u_2  & \cdots & 0 & 0 \\
\vdots & \vdots & \ddots & \vdots & \vdots \\
0 & 0 & \cdots & u_{k-1} & -x_2 \\
0 & 0 & \cdots & -x_1 & u_k \\
\end{vmatrix},\label{tridiagonal}\ee
where the $u_j$ are arbitrary coefficients, satisfy the recursion $D_k = u_k D_{k-1} - x_1x_2 D_{k-2}$ and are therefore polynomials in $x_1x_2$.
\item Combining the observations 
in the last two items, we see that the determinant is of the form
$$\Delta(x_1,x_2,y_1,y_2) = -x_1^q - x_2^q - y_1^q - y_2^q + R(x_1x_2,y_1y_2)$$
for some polynomial $R$. If we set $x_1=x_2=y_1=y_2 =0$, the determinant evaluates to $1$, so we can write $R(x_1x_2,y_1y_2) = 1 + V(x_1x_2,y_1y_2)$, where the constant coefficient of $V$ is zero.
It remains to show that $V$ is symmetric and has only real coefficients.

\medskip

If we exchange $y_1$ and $y_2$, then $\Delta(x_1,x_2,y_1,y_2) = 1  -x_1^q - x_2^q - y_1^q - y_2^q + V(x_1x_2,y_1y_2)$ obviously does not change. On the other hand, $c_j = 1 - \scq^jy_1-\scq^{-j}y_2$ becomes $1 - \scq^jy_2-\scq^{-j}y_1 = \overline{c_j}$. All other matrix entries are real and therefore equal to their conjugates (if we consider $x_1,x_2,y_1,y_2$ as real variables for the moment). This means that $\Delta$ is equal to its own conjugate, so it has only real coefficients.

\item Finally, we observe that $\Delta$ is an irreducible polynomial: suppose for contradiction that it can be factorised into two nonconstant factors. If we set $x_2 = y_2 = 0$, we obtain a factorisation of $1-x_1^q-y_1^q$, which is irreducible: its factorisation as a polynomial in $x_1$ is
$$1-x_1^q-y_1^q = - \prod_{j=0}^{q-1} (x_1-\scq^j (1-y_1^q)^{1/q}),$$
and since $1-y_1^q$ is not a power of a polynomial, there is no factorisation of $1-x_1^q-y_1^q$ into polynomials.

\medskip

Thus the factorisation of $\Delta$ becomes $(1 - x_1^q - y_1^q) \cdot 1$ for $x_2 = y_2 = 0$. So the first factor has total degree (at least) $q$, which is already the total degree of $\Delta$. This means that the second factor has to be constant, and we reach a contradiction.

\medskip

Therefore,
$$A(x_1,x_2,y_1,y_2,\scq) = \frac{U(x_1,x_2,y_1,y_2)}{1 - x_1^q - x_2^q - y_1^q - y_2^q + V(x_1x_2,y_1y_2)}$$
is in its lowest terms. In view of symmetry condition~\eqref{eq:refl2}, the function does not change when $x_1,x_2$ are replaced by $y_1,y_2$ respectively. Thus $V$ must be symmetric.
\end{itemize}

\section{ Hofstadter quantum mechanics  and Trace formula}

As stated in the introduction, $Z_{m_1,m_2,l_1,l_2}(\scq)$  happens to be 
of interest  for the quantum Hofstadter model \cite{Hofstadter},
thanks to the  mapping \eqref{map}  between the algebraic area generating function for closed walks of length $n$ evaluated at $\scq=e^{i\gamma}$ 
and the $n$-th moment of the quantum Hofstadter Hamiltonian $ {\rm Tr}\:H_{\gamma}^{n}$. 
Of particular interest is the commensurate flux   $\gamma=2\pi p/q$,
where $p$ and $q$ are relatively prime, so that  $\scq$ is a primitive $q$-th root of unity.

In the Landau gauge, the  Hofstadter  Hamiltonian is
$$H_{\gamma}=T_x+T_x^{-1}+T_y+T_y^{-1},$$
where the lattice hopping operators $T_x$ and $T_y$ obey  the  commutation relation
$$
T_x T_y = e^{-i\gamma} T_y T_x.
$$
They act on  a state   $\Psi_{m,n}$  at lattice site $\{m,n\}$ as follows: 
$$
T_x \Psi_{m, n} = \Psi _{m + 1,  n}, \quad T_y \Psi_{m, n}  = e^{i \gamma m} \Psi _{m , n + 1}.
$$
Using
translation invariance in the $y$ direction   one sets  $\Psi_{m,n}=e^{i n k_y }\Phi_m$ to get   the eigenenergy  Harper equation 
$$\Phi_{m+1}+\Phi_{m-1}+2\cos(k_y+\gamma m)\Phi_{m}=E\Phi_{m}.$$
In the commensurate case  
$
\gamma  = 2 \pi {p}/{q}
$
one has on the horizontal axis a periodic model with period $q$, so 
$\Phi_{m + q} = e^{i q k_x } \Phi_m$.
All this  amounts to  the $q\times q$ matrix $m_{p/q}(E,k_x,k_y)$  acting with zero output on the $q$-components eigenvector $\{ \Phi_0, \Phi_1,\ldots, \Phi_{q-1}\}$:
\be\label{matrixHof}\begin{pmatrix}
 2 \cos ({k_y})-E& 1 & 0 & \cdots & 0 & e^{-i {q k_x}} \\
1 & 2 \cos ({k_y}+\frac{2\pi p}{q})-E & 1 & \cdots & 0 & 0 \\
0 & 1 & () & \cdots & 0 & 0 \\
\vdots & \vdots & \vdots & \ddots & \vdots & \vdots \\
0 & 0 & 0  & \cdots & () & 1 \\
e^{i {q k_x}}  & 0 & 0  & \cdots & 1 & 2 \cos ({k_y}+(q-1)\frac{2\pi p}{q})-E  \\
\end{pmatrix} \cdot \begin{pmatrix} \Phi_0 \\ \Phi_1 \\ \Phi_2 \\ \vdots \\ \Phi_{q-2} \\ \Phi_{q-1} \end{pmatrix} = \begin{pmatrix}  0\\ 0 \\ 0 \\ \vdots \\ 0 \\ 0 \end{pmatrix},\ee
The $q$ eigenenergies $E_r(k_x,k_y)$ with $r=1, \ldots, q$ are   solutions of the secular equation 
$$\det(m_{p/q}(E,k_x,k_y))=0,$$
which, thanks to the identity 
$$\det(m_{p/q}(E,k_x,k_y))=\det(m_{p/q}(E,0,0))-2 (-1)^q (\cos(q k_x)-1 + \cos(q k_y)-1),$$
can be rewritten \cite{Chambers} as
\begin{equation} \det(m_{p/q}(E,0,0))=2 (-1)^q (\cos(q k_x)-1 + \cos(qk_y)-1).\label{eigen}\end{equation}
The trace   is  defined as
 \be\label{trace}
{\rm Tr}\; H_{2\pi p/ q}^n = \frac{1}{q}\int_{-\pi}^{\pi}\int_{-\pi}^{\pi} \frac{{\rm d} k_x}{2\pi}\frac{{\rm d} k_y}{2\pi} \sum_{r= 1}^{q} E_r^n(k_x,k_y),
\ee 
where  one sums  over the $q$  eigenenergies $E_r(k_x,k_y)$ of the Hofstadter Hamiltonian at power $n$ and integrates over the quasimomenta  $k_x\in[-\pi,\pi]$ and $k_y\in[-\pi,\pi]$.
So  in \eqref{trace}  computing\footnote{For  earlier attempts to compute such traces see for example \cite{Alexios}.}   ${\rm Tr}\:H_{2\pi p/ q}^{n}$  amounts to 
\begin{itemize}
\item first evaluating the determinant of $m_{p/q}(E,0,0)$, a polynomial of degree $q$ in $E$.
\item next solving the secular equation \eqref{eigen} for the $q$ eigenenergies $E_r(k_x,k_y)$; it can be done  numerically and leads to the  Hofstadter butterfly when  the eigenenergies  are plotted  against $p/q$.
\item finally summing and integrating  to get the trace of the Hofstadter Hamiltonian at a power $n$.
\end{itemize}

Evaluating ${\rm Tr}\; H_{2\pi p/ q}^n$  in this way (i.e., from its definition \eqref{trace} in terms of the quantum eigenenergies) is clearly a  difficult task.
We are going  to  address this question not from the quantum Hofstadter side,   but rather, via the mapping \eqref{map}, from the lattice walks combinatorial side,  by   evaluating  $Z_n(\scq)=\sum_{m=0}^{n/2}Z_{m, m, {n\over 2}-m, {n\over 2}-m}(\scq)$ at $\scq=e^{2i\pi p/q}$. 

In the following, we will derive an expression for the generating function
$ \sum_{n \geq 0} Z_n(e^{2i\pi p/q}) z^n$
in terms of a complete elliptic integral, evaluated at a rational function. To this end, we first specialise the multivariate generating function~\eqref{A_defi} from the previous section, i.e.
$$
A(x_1,x_2,y_1,y_2,\scq) = \sum_{m_1,m_2,l_1,l_2 \geq 0} Z_{m_1,m_2,l_1,l_2}(\scq) x_1^{m_1}x_2^{m_2} y_1^{l_1} y_2^{l_2}.
$$
One observes that $\sum_{m=0}^{n/2}Z_{m, m, {n\over 2}-m, {n\over 2}-m}(\scq)$ can be obtained directly from $A(x_1,x_2,y_1,y_2,\scq)$ by setting $x_1\to z x, x_2\to z/x, y_1\to z y$ and $ y_2\to z/y$ and looking at the coefficient where the exponents of $x$ and $y$ are $0$ ( enforcing $m_1-m_2=0$ and $l_1-l_2=0$) and the exponent of $z$ is $n$ (which corresponds to the condition $m_1+m_2+l_1+l_2=n$). 

\medskip

Let us now look for the coefficient of $x^0y^0$ in $A(zx,z/x,zy,z/y,\scq)$ evaluated at $\scq$ a root of unity: we  already know from~\eqref{gen_shape} that the determinant $\Delta$ in the denominator has  the form $ \Delta(x_1,x_2,y_1,y_2) = 1-x_1^q-x_2^q-y_1^q-y_2^q+V(x_1x_2,y_1y_2)$, so it simplifies to
$$\Delta(zx,z/x,zy,z/y) = 1-z^q(x^q+x^{-q}+y^q+y^{-q})+V(z^2,z^2).$$
For $x=y=1$,  we get $\Delta(z,z,z,z) = 1 - 4z^q + V(z^2,z^2)$, which is the determinant of the $q\times q$ matrix
\be\label{matrixwalk}\begin{pmatrix}
 1-2z & -z & 0 & \cdots & 0 & -z \\
-z & 1-2z \cos (\frac{2\pi p}{q})  &-z & \cdots & 0 & 0 \\
0 & -z & ()  & \cdots & 0 & 0 \\
\vdots & \vdots & \vdots & \ddots & \vdots & \vdots \\
0 & 0 & 0  & \cdots & () & -z \\
-z & 0 & 0  & \cdots & -z & 1-2z \cos ((q-1)\frac{2\pi p}{q}) \\
\end{pmatrix}. \ee
Hence we define  
\be\nonumber b_{p/ q}(z)=\Delta(z,z,z,z)+4z^q=1+V(z^2,z^2),\ee
a polynomial of degree $2\lfloor q/2 \rfloor$,   with  coefficients $-a_{p/ q}(2i)$:
\begin{equation}\label{kreftbis}b_{p/ q}(z) =-\sum_{i=0}^{\lfloor q/2 \rfloor}a_{p/ q}(2i)z^{2i}. \end{equation}
Now $\Delta(zx,z/x,zy,z/y)$  rewrites as 
\be\Delta(zx,z/x,zy,z/y) = b_{p/ q}(z)-z^q(x^q+x^{-q}+y^q+y^{-q}).\label{A_denom_simplified}\ee

Next we focus on the numerator of $A(zx,z/x,zy,z/y,\scq)$. It follows from~\eqref{A_denom_simplified} that the expansion of $1/\Delta$  only contains powers of $x$ and $y$ whose exponents are multiples of $-q$ and $q$. On the other hand, in view of its definition as a determinant in~\eqref{Cramer},
the numerator only contains powers of $x$ and $y$ with exponents between $-(q-1)$ and $q-1$. Since we are only interested in the coefficient of $x^0y^0$, we can focus on this coefficient in the numerator as well, as the other terms will only give rise to monomials in the expansion of $A(zx,z/x,zy,z/y,\scq)$ where the exponents of $x$ and $y$ are not simultaneously multiples of $q$.

As it turns out, the coefficient of $x^0y^0$ in the numerator can also be expressed in terms of the polynomial $b_{p/q}$:
\begin{equation}\label{eq:det_simple}
[x^0y^0] \begin{vmatrix}
1  & -z/x & 0 & \cdots & 0 & -zx \\
1 & c_1  & -z/x & \cdots & 0 & 0 \\
1 &-zx & c_2  & \cdots & 0 & 0 \\
\vdots & \vdots & \vdots & \ddots & \vdots & \vdots \\
1 & 0 & 0  & \cdots & c_{q-2} & -z/x \\
1 & 0 & 0  & \cdots & -zx & c_{q-1} \\
\end{vmatrix} = b_{p/ q}(z) - \frac{z}{q} b'_{p/ q}(z).
\end{equation}
A proof of this identity is given in the appendix.

Putting everything together that has been established so far for the denominator and numerator of $A(zx,z/x,zy,z/y,\scq)$, we are left with
$$[x^0y^0] A(zx,z/x,zy,z/y,\scq) = 
[x^0y^0] \frac{b_{p/ q}(z) - \frac{z}{q} b'_{p/ q}(z)}{b_{p/ q}(z) - z^q(x^q+x^{-q}+y^q+y^{-q})}.$$
This can be expanded as 
\begin{align*}
\frac{b_{p/ q}(z) - \frac{z}{q} b'_{p/ q}(z)}{b_{p/ q}(z) - z^q(x^q+x^{-q}+y^q+y^{-q})} &= \Big( 1 - \frac{zb'_{p/ q}(z)}{qb_{p/ q}(z)}\Big)  \frac{1}{1 - \frac{z^q}{b_{p/ q}(z)}(x^q+x^{-q}+y^q+y^{-q})} \\
&= \Big( 1 - \frac{zb'_{p/ q}(z)}{qb_{p/ q}(z)}\Big) \sum_{k \geq 0} \Big( \frac{z^q}{b_{p/ q}(z)} \Big)^k (x^q+x^{-q}+y^q+y^{-q})^k.
\end{align*}
The coefficient of $x^0y^0$ in $(x^q+x^{-q}+y^q+y^{-q})^k$ is $\binom{k}{k/2}^2$ if $k$ is even and $0$ otherwise,
so that this procedure finally coalesces to   
\be\label{semi-final}
\sum_{n \geq 0} Z_n(e^{2i\pi p/q}) z^n =\Big( 1 - \frac{zb'_{p/ q}(z)}{qb_{p/ q}(z)}\Big) \sum_{k \geq 0} \binom{2k}{k}^2 \Big( \frac{z^q}{b_{p/ q}(z)} \Big)^{2k}.
\ee
The series $\sum_{k \geq 0} \binom{2k}{k}^2 x^k$ can be expressed as $\frac{2}{\pi} K(16x)$, where $K$ is the complete elliptic integral of the first kind:
$$
K(x) = \int_0^1 \frac{1}{\sqrt{(1-t^2)(1-xt^2)}}\,{\rm d}t,
$$
see e.g. \cite[p. 161]{olver}. To conclude, we have 
\begin{equation}\label{eq:K-eq}
\sum_{n \geq 0} Z_n(e^{2i\pi p/q}) z^n = \Big( 1 - \frac{zb'_{p/ q}(z)}{qb_{p/ q}(z)}\Big) \frac{2}{\pi} K \Big( \frac{16z^{2q}}{b_{p/ q}(z)^2} \Big).
\end{equation}

For example, when $p=1$, $q=8$ i.e. $\scq = \exp(2\pi i/8)$, we get
$$b_{1/ 8}(z) = 1 - 16z^2+(72-8\sqrt{2})z^4-(96-32\sqrt{2})z^6+4z^8$$
and
$$ \sum_{n \geq 0} Z_n(e^{2\pi i/8}) z^n  = 1 + 4 z^2 +  (28 + 4\sqrt{2}) z^4 +  (232 + 72 \sqrt{2}) z^6 + (2140 + 960 \sqrt{2}) z^8 + \cdots.$$

Finding the generating function \eqref{eq:K-eq} for the $Z_n(e^{2i\pi p/q})$'s, or equivalently for the  traces ${\rm Tr}\; H_{2\pi p/ q}^n  $,  narrows down to determining the polynomial $b_{p/q}(z)$, i.e., the  coefficients $a_{p/ q}(2i)$ as defined in \eqref{kreftbis}. But  these coefficients of $b_{p/q}(z)$ in the expansion of the determinant of the matrix \eqref{matrixwalk} are  in one-to-one correspondence with those  of the expansion of the determinant of the Hofstadter matrix $m_{p/q}(E,0,0)$ in \eqref{matrixHof}, since it is easy to see that
\be\label{sure}(-1)^q E^qb_{p/q}(1/E)
= \det(m_{p/q}(E,0,0))+4(-1)^q,\ee
so that the Hofstadter secular energy equation \eqref{eigen} becomes
$$
E^qb_{p/q}(1/E)=2(\cos(q k_x)+\cos(q k_y)).
$$
In view of \eqref{sure}, the  $a_{p/q}(2i)$'s  can be identified with the  Kreft  coefficients ${\rm  c}_{p/q}(2i)$  defined in \cite{Kreft} 
$$ \det(m_{p/q}(E,0,0))+4(-1)^q=\sum_{i={q\over 2}-[{q\over 2}]}^q{\rm  c}_{p/q}(2i)E^{2i} $$
i.e.
$$  a_{p/q}(2i)={\rm  c}_{p/q}(q-2i)(-1)^{q+1},$$
 and one ends up with
\begin{equation}\label{thea}\small{
a_{p/q}(2i)=(-1)^{i+1}\sum_{k_1=0}^{q-2i}\sum_{k_2=0}^{k_1}\ldots\sum_{k_{i}=0}^{k_{i-1}} 4\sin ^2\left(\frac{\pi  (k_1+2i-1) p}{q}\right)4\sin ^2\left(\frac{\pi  (k_2+2i-3) p}{q}\right)\ldots 4\sin ^2\left(\frac{\pi  (k_{i}+1) p}{q}\right)}
\end{equation}
(see the appendix for some remarks on these coefficients). 

The elliptic integral $K$ satisfies a second-order differential equation with polynomial coefficients, therefore so does the generating function $\sum_{n \geq 0} Z_n(e^{2i\pi p/q}) z^n $ (although it is generally a rather complicated differential equation), i.e., it is holonomic. For example, when $q=4$,  (i.e. $\scq=i$), we have
\begin{align*}
&\big(4096 z^{15}-14848 z^{13}+17920 z^{11}-9696 z^9+2720 z^7-412 z^5+32 z^3-z\big) (\sum_{n \geq 0} Z_n(i) z^n)'' \\
&+\big(20480 z^{14}-54784 z^{12}+52480 z^{10}-26464 z^8+7040 z^6-940 z^4+56 z^2-1\big) (\sum_{n \geq 0} Z_n(i) z^n)' \\
&+\big(16384 z^{13}-32768 z^{11}+23040 z^9-7168 z^7+2112 z^5-320 z^3+16 z\big) (\sum_{n \geq 0} Z_n(i) z^n) = 0.
\end{align*}
This means that its coefficients, i.e., the traces $\operatorname{Tr} H_{2\pi p/q}^n$, satisfy a linear recursion with polynomial coefficients. For example, when $q=4$, the linear recursion is
\begin{align}
n^2 Z_n(i) &= \big(4096 n^2-98304 n+589824\big) Z_{n-14}(i)+\big(-14848 n^2+316416 n-1691648\big) Z_{n-12}(i) \nonumber\\
&\qquad +\big(17920 n^2-323840 n+1469440\big) Z_{n-10}(i)+\big(-9696 n^2+138368 n-493568\big) Z_{n-8}(i) \nonumber\\
&\qquad +\big(2720 n^2-28320 n+74112\big) Z_{n-6}(i)+\big(-412 n^2+2768 n-4800\big) Z_{n-4}(i) \\
&\qquad +\big(32 n^2-104 n+96\big) Z_{n-2}(i).\label{linear_rec}
\end{align}

It is also possible to give an explicit sum formula for the $Z_n(e^{2 i \pi p/q})$'s (equivalently, the traces $\operatorname{Tr} H_{2\pi p/q}^n$) in terms of the Kreft coefficients $a(2i)$: for even $n > 0$,
\be
Z_n(e^{2 i \pi p/q})={\rm Tr} H_{2\pi p/q}^{n} = \frac{n}{q} \sum_{k \geq 0} \sum_{\substack{\ell_1,\ell_2,\ldots,\ell_{\lfloor q/2 \rfloor} \geq 0 \\ \ell_1 + 2\ell_2 + \cdots + \lfloor q/2 \rfloor \ell_{\lfloor q/2 \rfloor} = n/2 - kq}}
\frac{\binom{2k}{k}^2 \binom{\ell_1+\ell_2 + \cdots + \ell_{\lfloor q/2 \rfloor} + 2k}{\ell_1,\ell_2,\ldots,\ell_{\lfloor q/2 \rfloor},2k}}{\ell_1+\ell_2 + \cdots + \ell_{\lfloor q/2 \rfloor} + 2k} \prod_{j = 1}^{\lfloor q/2 \rfloor} a(2j)^{\ell_j},\label{sumformula}
\ee
see the appendix for a complete derivation. For small values of $q$, this simplifies quite considerably ({\bf see \cite{Alexios}}): for $q=2$, we have
$$Z_n(-1) = \sum_{0 \leq k \leq n/4} \binom{2k}{k}^2 \binom{n/2}{2k} 2^{n-4k}.$$
Likewise, we get the following formula for $q=3$:
$$Z_n(e^{\pm 2\pi i/3}) = \frac{2n}{3} \sum_{0 \leq k \leq n/6} \binom{2k}{k}^2 \binom{n/2-k}{2k} \frac{6^{n/2-3k}}{n-2k}.$$

 For a given $n$, when $q$ increases as well as $p$---having in mind the irrational limit where both $p$ and $q\to \infty$---the  number of summation indices in \eqref{sumformula} increases as $\lfloor q/2 \rfloor +1$. Clearly the physical results  for, say, $p/q=1/2$ and $p/q=0.5000001$ should be very close to each other, at least if $n$ is not too large (for very large $n$, the difference actually goes to infinity). And yet, the first generating function is  a $2$-dimensional sum  while the second is a $5\;000\;001$-dimensional sum. This situation is reminiscent, for example,  of the  density correlation of the Calogero model obtained in \cite{ha} where  multi-dimensional integrals of greatly varying size are obtained for close rational Calogero  couplings.

As a final remark, we note that the generating function in \eqref{eq:K-eq} is  identical to 
$$\int_{-4}^4  \rho_{p/q}(E){1\over 1-z E} {\rm d} E=\Big( 1 - \frac{zb'_{p/ q}(z)}{qb_{p/ q}(z)}\Big) \frac{2}{\pi} K \Big( \frac{16z^{2q}}{b_{p/ q}(z)^2} \Big),$$
 where $\rho_{p/q}(E)$ is the Hofstadter density of states:  
$$\rho_{p/q}(E)={1\over 2\pi^2 q}|(E^{q}b_{p/ q}(1/E))'|K\left(1-\big({E^{q}b_{p/ q}(1/E)\over 4}\big)^2\right)$$
when 
$|{E^{q}b_{p/ q}(1/E)\over 4}|<1$, and zero otherwise \cite{wannier}.

\section{Asymptotics}
From the representation~\eqref{eq:K-eq} of the generating function, namely
$$
F_{p/q}(z) = \sum_{n \geq 0} Z_n(e^{2i\pi p/q}) z^n = \Big( 1 - \frac{zb'_{p/ q}(z)}{qb_{p/ q}(z)}\Big) \frac{2}{\pi} K \Big( \frac{16z^{2q}}{b_{p/ q}(z)^2} \Big),
$$
the asymptotic behaviour of $Z_n(e^{2i\pi p/q})$ as $n \to \infty$ for fixed $p/q$ can be obtained by standard means. We note that the elliptic integral $K$ has a logarithmic singularity at $1$:
$$K(x) =  -\frac12 \log(1-x) + O(1)$$
as $x \to 1$. This logarithmic singularity carries over to singularities of $F_{p/q}(z)$ at the points where $b_{p/q}(z) = \pm 4z^q$. For example, if $p/q = 1/2$, we have  $b_{1/2}(z) = 1-4z^2$, which gives us singularities at $z = \pm 1/\sqrt{8}$. At these singularities, the asymptotic behaviour of $F_{1/2}(z)$ is given by
$$F_{1/2}(z) = - \frac{2}{\pi} \log \big(1-\sqrt{8}z\big) + O(1)$$
and
$$F_{1/2}(z) = - \frac{2}{\pi} \log \big(1+\sqrt{8}z\big) + O(1)$$
respectively. Application of the Flajolet-Odlyzko singularity analysis \cite{flajolet} directly yields
$$Z_n(-1) = [z^n] F_{1/2}(z) \begin{cases} \sim \frac{4}{\pi n} \cdot 8^{n/2} & n \text{ even,} \\ = 0 & n \text{ odd.}\end{cases}$$

Similarly, for $p/q = 1/3$, we have $b_{1/3}(z) = 1-6z^2$, the dominant singularities (those closest to the origin) of $F_{1/3}(z)$ are $z = \pm \frac{\sqrt{3}-1}{2}$. We have
$$F_{1/3}(z) = -\frac{2+\sqrt{3}}{\pi} \log \Big(1 - \frac{2z}{\sqrt{3}-1} \Big) + O(1)$$
as $z \to \frac{\sqrt{3}-1}{2}$ and an analogous asymptotic formula as $z \to - \frac{\sqrt{3}-1}{2}$. Singularity analysis gives us
$$Z_n(e^{2\pi i/3}) = [z^n] F_{1/3}(z) \begin{cases} \sim \frac{4+2\sqrt{3}}{\pi n} \cdot (1+\sqrt{3})^n & n \text{ even,} \\ = 0 & n \text{ odd.}\end{cases}$$

In the same way, one obtains an asymptotic formula of the form
$$Z_n(e^{2\pi i p/q}) = [z^n] F_{p/q}(z) \sim \frac{\beta}{n} \cdot \alpha^n$$
for arbitrary fixed $p/q$ and even $n$ as $n \to \infty$, where $\alpha$ and $\beta$ depend on $p/q$. The table lists a few further values:

\begin{center}
\begin{tabular}{l|cccc}
$\frac{p}{q}$ & $\frac14$ or $\frac34$ & $\frac15$ or $\frac45$ & $\frac25$ or $\frac35$ & $\frac16$ or $\frac56$ \\
\hline
$\alpha$ & $\sqrt{8}$ & $\frac{1+\sqrt{5}+\sqrt{70+2\sqrt{5}}}{4}$ & $\frac{3+\sqrt{5}}{2}$ & $\sqrt{5+\sqrt{21}}$ \\
$\beta$ & $\frac{16}{\pi}$ & $\frac{19+11\sqrt{5}+\sqrt{670+298\sqrt{5}}}{2\pi}$ & $\frac{7+3\sqrt{5}}{\pi}$ & $\frac{56+12\sqrt{21}}{\pi}$
\end{tabular}
\end{center}

\section{Conclusion} 

Making use of the connection to the algebraic area of lattice walks, we are able to compute the generating function of the traces ${\rm Tr}\:H_{2\pi p/q}^{n}$ of the Hofstadter Hamiltonian  for any fixed rational number $p/q$---see~\eqref{eq:K-eq}. From this, one can derive both recursive (see~\eqref{linear_rec}) and explicit formulas (see~\eqref{sumformula}) for these traces. The building blocks of \eqref{sumformula}  are the Kreft coefficients in (\ref{thea}) (see also  section 6.2. in the Appendix).  When $q$ is small, these formulas turn out to be quite simple. The generating function can also be used to study the asymptotic behaviour as $n \to \infty$, which follows a law of the form ${\rm Tr}\:H_{2\pi p/q} \sim \frac{\beta}{n} \cdot \alpha^n$, where $\alpha$ and $\beta$ are constants depending on $p/q$.

\section*{Acknowledgements}

S.O. would like to thank Alain Comtet and Alexios Polychronakos for interesting conversations. He would also like to thank the AIMS SA Centre in Muizenberg, where part of this work was done, for its hospitality. S.W.'s research was supported by the National Research Foundation of South Africa, grant number 96236.

\section{Appendix} 

\subsection{Proof of~\eqref{eq:det_simple}}

\begin{enumerate}
\item Consider the bivariate determinant
$$B(y,z) = \begin{vmatrix}
c_0 & -z & 0 & \cdots & 0 & -z \\
 -z & c_1 & -z & \cdots & 0 & 0 \\
 0 & -z & c_2 & \cdots & 0 & 0 \\
\vdots & \vdots & \vdots & \ddots & \vdots & \vdots \\
0 & 0 & 0 & \cdots & c_{q-2} & -z \\
-z & 0 & 0 & \cdots & -z & c_{q-1}
\end{vmatrix},$$
with $c_k = 1 - z(\scq^ky - \scq^{-k}y^{-1})$. It is a priori clear that the determinant $B(y,z)$ can only contain the powers $y^{-q},y^{-q+1},\ldots,y^{-1},y^0,y^1,\ldots,y^{q-1},y^q$ and $z^0,z^1,\ldots,z^q$ respectively.  If $y$ is replaced by $\scq y$, then $c_k$ becomes $c_{k+1}$ (and $c_{q-1}$ becomes $c_0$), so this amounts to a cyclic permutation of the diagonal entries. Since this does not change the value of the determinant, we have $B(\scq y,z) = B(y,z)$. The only powers of $y$ that stay invariant under the transformation $y \mapsto \scq y$ are $y^{-q}$, $y^0$ and $y^q$, so these are the only powers that actually occur in $B(y,z)$. Moreover, making use of the fact that $y$ and $y^{-1}$ only occur in the diagonal entries, one easily finds that
$$[y^q] B(y,z) = [y^q] \prod_{k=0}^{q-1} c_k = (-1)^q \scq^{q(q-1)/2}z^q = \pm z^q,$$
and likewise $[y^{-q}] B(y,z) = \pm z^q$. Since the term involving $z^q$ cancels in $B(y,z) - \frac{z}{q} \frac{\partial}{\partial z} B(y,z)$, this means that $B(y,z) - \frac{z}{q} \frac{\partial}{\partial z} B(y,z)$ does not contain $y$ at all (i.e., it is independent of $y$). Thus we can also write
$$b_{p/q}(z) - \frac{z}{q} b'_{p/q}(z) = B(1,z) - \frac{z}{q} \frac{\partial}{\partial z} B(1,z) = [y^0]\Big( B(y,z) - \frac{z}{q} \frac{\partial}{\partial z} B(y,z) \Big).$$
\item Next, we apply Jacobi's formula for the derivative of a determinant \cite[Part Three, Section 8.3]{magnus} to the matrix
$$M(y,z) = \begin{pmatrix}
c_0 & -z & 0 & \cdots & 0 & -z \\
 -z & c_1 & -z & \cdots & 0 & 0 \\
 0 & -z & c_2 & \cdots & 0 & 0 \\
\vdots & \vdots & \vdots & \ddots & \vdots & \vdots \\
0 & 0 & 0 & \cdots & c_{q-2} & -z \\
-z & 0 & 0 & \cdots & -z & c_{q-1}
\end{pmatrix}$$
to obtain
$$\frac{\partial}{\partial z} B(y,z) = \frac{\partial}{\partial z} \det(M(y,z)) = \mtr \Big( \adj(M(y,z)) \frac{\partial M(y,z)}{\partial z} \Big).$$
This gives us, with $I_q$ denoting the $q \times q$ identity matrix,
\begin{align*}
B(y,z) - \frac{z}{q} \frac{\partial}{\partial z} B(y,z) &= \det(M(y,z)) - \frac{z}{q} \mtr \Big( \adj(M(y,z)) \frac{\partial M(y,z)}{\partial z} \Big) \\
&= \frac{1}{q} \mtr \Big(\det(M(y,z)) I_q \Big) - \frac{z}{q} \mtr \Big( \adj(M(y,z)) \frac{\partial M(y,z)}{\partial z} \Big) \\
&= \frac{1}{q} \mtr \Big(\adj(M(y,z)) M(y,z)\Big) - \frac{z}{q} \mtr \Big( \adj(M(y,z)) \frac{\partial M(y,z)}{\partial z} \Big) \\
&= \frac{1}{q} \mtr \Big(\adj(M(y,z)) \Big( M(y,z) - \frac{z}{q} \frac{\partial M(y,z)}{\partial z} \Big) \Big) \\
&= \frac{1}{q} \mtr \Big(\adj(M(y,z)) I_q \Big) \\
&= \frac1q \mtr \Big( \adj(M(y,z)) \Big).
\end{align*}
\item The trace in this formula is the sum of $q$ minors of $M(y,z)$, namely the determinants of the matrices obtained from $M(y,z)$ by removing a row and the corresponding column. Let these matrices be denoted $M_1(y,z), M_2(y,z), \ldots, M_q(y,z)$. We note that $M_k(y,z)$ can be obtained from $M_1(y,z)$ by the substitution $y \mapsto \scq^{k-1}y$ and cyclic permutation of rows and columns. It follows that 
$$[y^0] M_1(y,z) = [y^0] M_2(y,z) = \cdots = [y^0] M_q(y,z),$$
so
\begin{align*}
b_{p/q}(z) - \frac{z}{q} b'_{p/q}(z) &= [y^0]\Big( B(y,z) - \frac{z}{q} \frac{\partial}{\partial z} B(y,z) \Big) \\
&= [y^0] \frac1q \Big(M_1(y,z) + M_2(y,z) + \cdots + M_q(y,z) \Big) \\
&= [y^0] M_1(y,z).
\end{align*}
\item On the other hand, if we expand the determinant 
$$\begin{vmatrix}
1  & -z/x & 0 & \cdots & 0 & -zx \\
1 & c_1  & -z/x & \cdots & 0 & 0 \\
1 &-zx & c_2  & \cdots & 0 & 0 \\
\vdots & \vdots & \vdots & \ddots & \vdots & \vdots \\
1 & 0 & 0  & \cdots & c_{q-2} & -z/x \\
1 & 0 & 0  & \cdots & -zx & c_{q-1}
\end{vmatrix}$$
first with respect to the first column, then with respect to the first row, we obtain
\begin{equation}\label{eq:expansion}
\begin{split}
\begin{vmatrix}
c_1  & -z/x & \cdots & 0 & 0 \\
-zx & c_2  & \cdots & 0 & 0 \\
\vdots & \vdots  & \ddots & \vdots & \vdots \\
0 & 0  & \cdots & c_{q-2} & -z/x \\
0 & 0  & \cdots & -zx & c_{q-1}
\end{vmatrix} + \sum_{r=1}^{q-1} \frac{z^r}{x^r} \begin{vmatrix}
c_{r+1}  & -z/x & \cdots & 0 & 0 \\
-zx & c_{r+2}  & \cdots & 0 & 0 \\
\vdots & \vdots  & \ddots & \vdots & \vdots \\
0 & 0  & \cdots & c_{q-2} & -z/x \\
0 & 0  & \cdots & -zx & c_{q-1}
\end{vmatrix} \\
+ 
\sum_{r=1}^{q-1} z^r x^r \begin{vmatrix}
c_1  & -z/x & \cdots & 0 & 0 \\
-zx & c_2  & \cdots & 0 & 0 \\
\vdots & \vdots &  \ddots & \vdots & \vdots \\
0 & 0  & \cdots & c_{q-r-2} & -z/x \\
0 & 0  & \cdots & -zx & c_{q-r-1}
\end{vmatrix},
\end{split}
\end{equation}
where an empty determinant is to be interpreted as $1$. Now we make use of the prior observation (see \eqref{tridiagonal}) that determinants of the form
$$D_k = \begin{vmatrix}
u_1  & -x_2 & \cdots & 0 & 0 \\
-x_1 & u_2  & \cdots & 0 & 0 \\
\vdots & \vdots & \ddots & \vdots & \vdots \\
0 & 0 & \cdots & u_{k-1} & -x_2 \\
0 & 0 & \cdots & -x_1 & u_k \\
\end{vmatrix},$$
where the $u_j$ are arbitrary coefficients, satisfy the recursion $D_k = u_k D_{k-1} - x_1x_2 D_{k-2}$. Here, this implies that all the determinants in~\eqref{eq:expansion} are actually independent of $x$, so
\begin{align*}
[x^0 y^0] \begin{vmatrix}
1  & -z/x & 0 & \cdots & 0 & -zx \\
1 & c_1  & -z/x & \cdots & 0 & 0 \\
1 &-zx & c_2  & \cdots & 0 & 0 \\
\vdots & \vdots & \vdots & \ddots & \vdots & \vdots \\
1 & 0 & 0  & \cdots & c_{q-2} & -z/x \\
1 & 0 & 0  & \cdots & -zx & c_{q-1}
\end{vmatrix} &= [x^0y^0] \begin{vmatrix}
c_1  & -z/x & \cdots & 0 & 0 \\
-zx & c_2  & \cdots & 0 & 0 \\
\vdots & \vdots & \ddots & \vdots & \vdots \\
0 & 0  & \cdots & c_{q-2} & -z/x \\
0 & 0  & \cdots & -zx & c_{q-1}
\end{vmatrix} \\
&= [y^0] \begin{vmatrix}
c_1  & -z & \cdots & 0 & 0 \\
-z & c_2  & \cdots & 0 & 0 \\
\vdots & \vdots  & \ddots & \vdots & \vdots \\
0 & 0  & \cdots & c_{q-2} & -z \\
0 & 0  & \cdots & -z & c_{q-1}
\end{vmatrix} \\
&= [y^0] M_1(y,z) \\
&= b_{p/q}(z) - \frac{z}{q} b'_{p/q}(z),
\end{align*}
completing the proof of~\eqref{eq:det_simple}.
\end{enumerate}

\subsection{The Kreft coefficients}

In \eqref{thea} the coefficients $a_{p/q}(2i)$   are properly defined when $q\ge 2i$:
one has 
\begin{align*}
a_{p/q}(0) &=-1, \\
a_{p/q}(2) &=\sum_{i_1=0}^{q-2}4\sin ^2\left(\frac{\pi  (i_{1}+1) p}{q}\right), \\
a_{p/q}(4) &=-\sum_{i_1=0}^{q-4}\sum_{i_2=0}^{i_1} 4\sin ^2\left(\frac{\pi  (i_1+3) p}{q}\right) 4\sin ^2\left(\frac{\pi  (i_{2}+1) p}{q}\right), \\
a_{p/q}(6) &=\sum_{i_1=0}^{q-6}\sum_{i_2=0}^{i_1}\sum_{i_3=0}^{i_2} 4\sin ^2\left(\frac{\pi  (i_1+5) p}{q}\right) 4\sin ^2\left(\frac{\pi  (i_{2}+3) p}{q}\right)4\sin ^2\left(\frac{\pi  (i_{3}+1) p}{q}\right),
\end{align*}
etc. One obtains in this way that
    \begin{align*}
a_{p/q}(2)&=2 q,\\
a_{p/q}(4)&= -2q^2 + 7q + 2q \cos \left(\frac{2 \pi  p}{q}\right),\\
a_{p/q}(6)&=\frac{4q^3-42q^2+116q}{3} + (-4q^2+24q) \cos \left(\frac{2 \pi p}{q}\right) + 4q \cos \left(\frac{4 \pi p}{q}\right),\\
a_{p/q}(8)&=\frac{-4q^4+84q^3-617q^2+1617q}{6} + (4q^3-62q^2+252q)  \cos \left(\frac{2 \pi  p}{q}\right) \\
&\quad+ (-9q^2+77q) \cos \left(\frac{4 \pi  p}{q}\right) + 12q \cos \left(\frac{6 \pi  p}{q}\right) + 2q \cos \left(\frac{8 \pi p}{q}\right),\\
   a_{p/q}(10)&=\frac{4q^5-140q^4+1925q^3-12505q^2+32916q}{15} + \frac{-8q^4+228q^3-2260q^2+7896q}{3} \cos \left(\frac{2 \pi p}{q}\right) \\
&\quad+ (10q^3-206q^2+1108q) \cos \left(\frac{4 \pi p}{q}\right) + (-28q^2+312q)\cos \left(\frac{6 \pi p}{q}\right) \\
&\quad+ (-4q^2+84q) \cos \left(\frac{8 \pi p}{q}\right) + 16q \cos \left(\frac{10 \pi p}{q}\right) + 4q \cos \left(\frac{12 \pi p}{q}\right),
\end{align*} 
etc.

\subsection{Proof of the trace formula \eqref{sumformula}.} 

Let us return to the representation~\eqref{semi-final}:
$$\sum_{n \geq 0} Z_n(e^{2i\pi p/q}) z^n =\Big( 1 - \frac{zb'_{p/ q}(z)}{qb_{p/ q}(z)}\Big) \sum_{k \geq 0} \binom{2k}{k}^2 \Big( \frac{z^q}{b_{p/ q}(z)} \Big)^{2k}.$$
We notice first that
$$ 1 - \frac{zb'_{p/ q}(z)}{qb_{p/ q}(z)} = \frac{z}{q} \cdot \frac{\frac{d}{dz} \frac{z^q}{b_{p/ q}(z)}}{\frac{z^q}{b_{p/ q}(z)}},$$
so this can also be rewritten as
\begin{align*}
\sum_{n \geq 0} Z_n(e^{2i\pi p/q}) z^n &= \frac{z}{q} \sum_{k \geq 0} \binom{2k}{k}^2 \Big( \frac{z^q}{b_{p/ q}(z)} \Big)^{2k-1}\frac{d}{dz} \frac{z^q}{b_{p/ q}(z)} \\
&= \frac{z}{q} \frac{d}{dz} \bigg( \log \Big( \frac{z^q}{b_{p/ q}(z)} \Big) + \sum_{k \geq 1} \binom{2k}{k}^2 \frac{1}{2k} \Big( \frac{z^q}{b_{p/ q}(z)} \Big)^{2k} \bigg).
\end{align*}
It follows that
$$Z_n(e^{2i\pi p/q}) = \frac{n}{q} [z^n] \bigg( \log \Big( \frac{z^q}{b_{p/ q}(z)} \Big) + \sum_{k \geq 1} \binom{2k}{k}^2 \frac{1}{2k} \Big( \frac{z^q}{b_{p/ q}(z)} \Big)^{2k} \bigg).$$
Now set $s_{p/q}(z) = 1 - b_{p/q}(z) = \sum_{i=1}^{\lfloor q/2 \rfloor} a(2i) z^{2i}$ and let us expand $-\log(b_{p/ q}(z)) = - \log(1-s_{p/q}(z))$ and $b_{p/ q}(z)^{-2k} = (1-s_{p/q}(z))^{-2k}$ into series. Note also that $\log(z^q) = q \log z$ does not contribute to the coefficient of $z^n$. We obtain
\begin{align*}
Z_n(e^{2i\pi p/q}) &= \frac{n}{q} [z^n] \bigg( -\log(1-s_{p/q}(z)) + \sum_{k \geq 1} \binom{2k}{k}^2 \frac{z^{2kq}}{2k} (1-s_{p/q}(z))^{2k} \bigg) \\
&= \frac{n}{q} [z^n] \bigg( \sum_{m \geq 1} \frac{s_{p/q}(z)^m}{m} + \sum_{k \geq 1} \binom{2k}{k}^2 \frac{z^{2kq}}{2k} \sum_{m \geq 0} \binom{m+2k-1}{2k-1} s_{p/q}(z)^m \bigg).
\end{align*}
The two sums can be combined to a single one:
\be\label{last_step}
Z_n(e^{2i\pi p/q}) = \frac{n}{q} [z^n] \sum_{\substack{k,m \geq 0 \\ (k,m) \neq (0,0)}}  \frac{\binom{2k}{k}^2 \binom{m+2k}{2k}}{m+2k} z^{2kq} s_{p/q}(z)^m
= \frac{n}{q} \sum_{\substack{k,m \geq 0 \\ (k,m) \neq (0,0)}}  \frac{\binom{2k}{k}^2 \binom{m+2k}{2k}}{m+2k} [z^{n-2kq}]s_{p/q}(z)^m.
\ee
When $s_{p/q}(z)^m = \big( \sum_{i=1}^{\lfloor q/2 \rfloor} a(2i) z^{2i} \big)^m$ is expanded, the resulting terms are of the form
$$\binom{m}{\ell_1,\ell_2,\ldots,\ell_{\lfloor q/2 \rfloor}} z^{2(\ell_1+2\ell_2+\cdots+\lfloor q/2 \rfloor\ell_{\lfloor q/2 \rfloor})} \prod_{j = 1}^{\lfloor q/2 \rfloor} a(2j)^{\ell_j},$$
where $\ell_1 + \ell_2 + \cdots + \ell_{\lfloor q/2 \rfloor} = m$. Since we are taking the coefficient of $z^{n-2kq}$, the $\ell_j$ have to satisfy
$$\ell_1 + 2\ell_2 + \cdots + \lfloor q/2 \rfloor \ell_{\lfloor q/2 \rfloor} = n/2 - kq.$$
Putting everything together, we arrive at~\eqref{sumformula}. Note that $k=m=0$ is impossible for $n > 0$, so the restriction in the sum~\eqref{last_step} is actually immaterial.

\end{document}